\documentclass[sn-mathphys-num]{sn-jnl}

\usepackage{graphicx}%
\usepackage{multirow}%
\usepackage{amsmath,amssymb,amsfonts}%
\usepackage{amsthm}%
\usepackage{mathrsfs}%
\usepackage[title]{appendix}%
\usepackage{xcolor}%
\usepackage{textcomp}%
\usepackage{manyfoot}%
\usepackage{booktabs}%
\usepackage{listings}%

\def\bea{\begin{eqnarray}}
\def\eea{\end{eqnarray}}
\def\be{\begin{equation}}
\def\ee{\end{equation}}

\newcommand{\Pminus}{{\cal P}^-}
\newcommand{\Pfree}{\Pminus_0}
\newcommand{\Pint}{\Pminus_{\rm int}}
\newcommand{\Pvac}{P_{\rm vac}^-}
\newcommand{\vac}{|{\rm vac}\rangle}
\newcommand{\deps}{\delta_\epsilon}

\begin{document}

\title[Zero Modes on the Light Front]{Zero Modes on the Light Front}


\author[1]{\fnm{S.S.} \sur{Chabysheva}}\email{schabysheva@uidaho.edu}

\author*[1,2]{\fnm{J.R.} \sur{Hiller}}\email{jhiller@d.umn.edu}
\equalcont{These authors contributed equally to this work.}

\affil*[1]{\orgdiv{Department of Physics}, \orgname{University of Idaho}, \orgaddress{\street{875 Perimeter Drive}, \city{Moscow}, \postcode{83844}, \state{Idaho}, \country{USA}}}

\affil[2]{\orgdiv{Department of Physics and Astronomy}, \orgname{University of Minnesota-Duluth}, \orgaddress{\street{1023 University Drive}, \city{Duluth}, \postcode{55812}, \state{Minnesota}, \country{USA}}}


\abstract{Modes with zero longitudinal light-front momentum (zero modes) do have roles to play 
in the analysis of light-front field theories.  These range from improvements in convergence 
for numerical calculations to implications for the light-front vacuum and beyond to fundamental 
issues in the connection with equal-time quantization.  In particular, the discrepancy in values 
of the critical coupling for $\phi^4_{1+1}$ theory, between equal-time and light-front quantizations, 
would appear to be resolvable with the proper treatment of zero modes and near-zero modes.  We provide 
a survey of these issues and point to open questions.
}

\keywords{light front, zero modes, vacuum structure}



\maketitle

\section{Introduction}\label{sec:intro}

Almost all light-front (LF) calculations begin with the assumption
of a trivial vacuum, identified with the Fock vacuum.\footnote{See \protect\cite{Dirac}
for the initial formulation of light-front coordinates and 
see ~\protect\cite{BPP,Carbonell,Miller,Heinzl,Burkardt,Hiller}
for reviews of light-front methods.}  
This is because the LF longitudinal momentum of a massive particle
$p^-=\sqrt{m^2+p_z^2+p_\perp^2}+p_z$ 
is always greater than zero and seemingly incapable of 
contributing to a vacuum with zero momentum.  In other
words, modes of zero longitudinal momentum, zero modes,
need to be present if the vacuum is to be nontrivial.  
So, one frequently sees a calculation
begin with the assumption that zero modes will be neglected.
Depending on the purpose of the particular calculation,
this assumption may or may not create difficulties.

Certainly there must be the concern that something is
missed in the neglect of zero modes and the assumed
triviality of the vacuum.  Many phenomena, such as
condensates~\cite{ChiralSymmetry}, 
spontaneous symmetry breaking~\cite{WernerHeinzl,Robertson,BenderPinsky,PinskyvdSHiller,Tsujimaru,MartinovicVary},
and zero-mode contributions to twist-3 GPDs~\cite{GPDs} are
nominally associated with vacuum structure.  A trivial
LF vacuum makes these difficult to explain and understand.

One example where zero modes clearly play a role is
in the calculation of the critical coupling for two-dimensional
$\phi^4$ theory~\cite{BCH,HiResVary,Katz1,LeeSalwen,Sugihara,SchaichLoinaz,Bosetti,%
Milsted,RychkovVitale,Katz2,Katz3}.\footnote{See \protect\cite{Tadpoles}
for additional references.}
Calculations done with equal-time (ET) and LF
quantizations disagree, with values separated by several $\sigma$
in the estimates of numerical errors.  As evidence, we include
Table~\ref{tab:phi4crit} with recent values obtained.  The LF
results are significantly smaller in value.
However, extrapolations from ET quantization to the LF~\cite{Hornbostel,Ji,ETtoLF}
show that the LF result for the critical coupling
should instead be consistent with the ET value.

\begin{table}[ht]
\caption{\label{tab:phi4crit}
Tabulation of critical coupling values for $\phi^4$ theory
with the coupling $\lambda$ defined by the 
interaction Lagrangian ${\cal L}_{\rm int}=-\lambda\phi^4/4!$.
The first two values were computed
in light-front quantization and the remainder in equal-time
quantization; the results are inconsistent unless the
different mass renormalizations are taken into account.}
\begin{tabular}{lll}
\toprule
Method &  $g_c\equiv\lambda_c/(24\mu^2)$ & Reported by \\
\midrule
Light-front symmetric polynomials & $1.1\pm0.03$ & Burkardt {\em et al.}~\protect\cite{BCH} \\
DLCQ  & $0.94\pm0.01$ & Vary {\em et al.}~\protect\cite{HiResVary} \\
Conformal light-front truncation & $0.96\pm0.02$ & Anand {\em et al.}~\protect\cite{Katz1} \\
Quasi-sparse eigenvector & 2.5 & 
     Lee \& Salwen~\protect\cite{LeeSalwen} \\
Density matrix renormalization group & 2.4954(4) & 
     Sugihara~\protect\cite{Sugihara} \\
Lattice Monte Carlo & 2.70$\left\{\begin{array}{l} +0.025 \\ -0.013\end{array}\right.$ & 
     Schaich \& Loinaz~\protect\cite{SchaichLoinaz} \\
                    & $2.79\pm0.02$ & Bosetti {\em et al.}~\protect\cite{Bosetti} \\
Uniform matrix product & 2.766(5) & 
     Milsted {\em et al.}~\protect\cite{Milsted} \\
Renormalized Hamiltonian truncation & 2.97(14) & 
     Rychkov \& Vitale~\protect\cite{RychkovVitale} \\
\botrule
\end{tabular}
\end{table}

As emphasized by Burkardt~\cite{MuchAdo}, and 
checked explicitly in \cite{BCH}, the difference is
due to the absence of tadpole contributions in the LF
calculations.  Tadpole contributions involve a coupling of
vacuum-to-vacuum transitions to a propagating constituent
that alters the mass renormalization.  With no zero modes
and only a trivial vacuum, LF calculations have no
vacuum-to-vacuum transitions; intermediate states have
nonzero momentum and cannot couple to the vacuum.  However,
the necessary tadpole correction is computable by taking
expectation values of powers of the field~\cite{MuchAdo}.
Under these circumstances, the bare masses in LF and
ET quantizations are related by
\be
\mu^2_{\rm LF}=\mu^2_{\rm ET}+\lambda\left[\langle0|\frac{\phi^2}{2}|0\rangle
                       -\langle0|\frac{\phi^2}{2}|0\rangle_{\rm free}\right].
\ee
The matrix element labeled `free' is computed with zero coupling.
With different mass scales present, the dimensionless coupling
$g\sim \lambda/\mu^2$ also differs.  Within this framework,
the difference in ET and LF couplings implies that values
of the dimensionless critical coupling will also be different.
Calculations done with this approach~\cite{BCH} indicate
that the difference between ET and LF results can be
explained in this way.

However, this is not without some difficulty in the extrapolation
to the critical value.  The computation of the expectation value
$\langle\phi^2\rangle$ used a spectral decomposition~\cite{BCH}.
Near the critical coupling the mass spectrum was poorly approximated
by the numerical algorithm, hence the need for extrapolation from
weaker couplings.  Fitzpatrick {\em et al}.~\cite{Katz3} interpret
this as a fundamental difficulty with the nonperturbative mass
renormalization itself and propose an alternative renormalization,
though there may be numerical artifacts there as well.  In any 
case, the approach reviewed here is done wholly within LF
quantization, motivated by the need to include tadpole 
contributions.\footnote{An alternative approach to the incusion 
of zero-mode effects is explained in~\cite{Fitzpatrick}, in the 
context of a LF Ising model where zero modes are integrated out to
obtain an effective interaction.}

This shows that the piece missing from LF calculations is
the possibility of vacuum-to-vacuum transitions~\cite{Tadpoles}.  
Although the positivity of longitudinal momentum is what is used
to argue for their absence, it is the neglect of terms
with only annihilation or only creation operators in the
LF Hamiltonian that truly eliminates the possibility, because
transitions between the trivial vacuum and intermediate
states cannot then take place.

That such terms should be excluded contradicts the fact
that matrix elements of only annihilation or only
creation operators between Fock states can be nonzero,
depending on the endpoint behavior of the wave functions.
The endpoints themselves are zero modes and only a set
of measure zero with respect to the integration measure
for the longitudinal momentum.  It is in the limit to the zero mode
that the momentum integrals obtain contributions.  This is 
consistent with Yamawaki's view~\cite{Tsujimaru} of zero
modes as accumulation points. The essential point is 
that Hamiltonian terms with operators in the integrand
are not just functions, and the momentum conserving delta
function that enforces the net momentum of zero
should not be invoked until the operators are used to
obtain an ordinary integrand.

To see that such matrix elements can be nonzero, consider
a scalar theory where, as is typical, the interaction
Hamiltonian contains a factor of $1/\sqrt{p_1p_2p_3}$
with $p_i$ an LF momentum at the vertex.  The equation
for a Fock-state wave function for $n$ constituents
coupled to such an interaction takes the form
\bea
\lefteqn{
\sum_i^n \frac{\mu^2}{p_i^+}\psi_n(p_1^+,\ldots,p_n^+)}&& \\
&&+\frac{1}{\sqrt{p_1^+p_2^+p_3^+\ldots}}\times\mbox{contributions from other Fock sectors}=\frac{M^2}{P^+}\psi_n.
\nonumber
\eea
The requirement of symmetry for identical bosons then implies
that the small-momentum behavior of the wave function is
\be  \label{eq:behavior}
\psi_n\sim \frac{1}{\sqrt{\prod_i^n p_i^+}\sum_i^n\frac{1}{p_i^+}}.
\ee
For wave functions with such behavior, a matrix element in $\phi^4$ theory
for the all-creation term
\be
\int\frac{\prod_i^4 dp^+_i}{\sqrt{\prod_i^4 p^+_i}}\delta(\sum_i^4 p^+_i)\prod_i^4 a^\dagger(p^+_i)
\ee
between Fock states with $n$ and $n+4$ constituents is
\bea
\lefteqn{\langle \psi_{n+4}|\int \frac{\prod_i^4 dp_i^+}{\sqrt{\prod_i^4 p_i^+}}
  \delta(\sum_i^4 p_i^+)\prod_i^4 a^\dagger(p_i^+)|\psi_n\rangle}&& \nonumber \\
&&\sim \int \frac{\prod_i^{n+4} dp_i^+}{\prod_i^{n+4} p_i^+}
           \frac{\delta(\sum_i^4 p_{n+i}^+)}
              {\left(\sum_i^{n+4}\frac{1}{p_i^+}\right)\left(\sum_i^n\frac{1}{p_i^+}\right)}
              \delta(P^+-\sum_i^n p_i^+),
\eea
where we have included the constraint of the total LF momentum being $P^+$.  This 
integral is finite, as can be seen by introducing a total momentum 
$Q\equiv\sum_i^4 p_{n+i}^+$ and relative momenta, $x_i\equiv p_{n+i}^+/Q$ for 
the created particles, as introduced in~\cite{Tadpoles}.  The Jacobian for the transformation is
specified by
\be
\prod_{i=n+1}^{n+4} dp_i^+=Q^3dQ\prod_i^4 dx_i\delta(1-\sum_i^4x_i),
\ee
and the matrix element becomes
\bea
\lefteqn{\langle \psi_{n+4}|\int \frac{\prod_i^4 dp_i^+}{\sqrt{\prod_i^4 p_i^+}}
  \delta(\sum_i^4 p_i^+)\prod_i^4 a^\dagger(p_i^+)|\psi_n\rangle} & & \\
&& \sim \int_0^\infty dQ \delta(Q)  \int \frac{\prod_i^4 dx_i \delta(1-\sum_i^4 x_i)}
                                      {\left(\prod_i^4 x_i\right)\left(\sum_i^4\frac{1}{x_i}\right)}
   \int \frac{\prod_i^n dp_i^+ \delta(P^+-\sum_i^n p_i^+)}
      {\left(\prod_i^n p_i^+\right)\left(\sum_i^n\frac{1}{p_i^+}\right)},
   \nonumber
\eea
where we have used $p_{n+i}^+=Qx_i$ with $Q\rightarrow0$ to replace
$\sum_i^{n+4}\frac{1}{p_i^+}$ with $\frac{1}{Q}\sum_i^4\frac{1}{x_i}$.
This expression is finite and nonzero, with\footnote{The integral 
$\int_0^\infty\delta(Q)dQ$ is equal to 1/2 because 
$\int_{-\infty}^\infty\delta(Q)dQ=1$ and $\delta(Q)$ is 
an even function.} $\int_0^\infty dQ \delta(Q)=\frac12$.

The vacuum-to-vacuum transitions that admit tadpole
contributions also allow vacuum bubbles.  In a 
perturbative calculation, bubbles can be eliminated
by hand; one simply neglects such graphs.  However,
in a nonperturbative calculation, explicit subtraction
is usually not possible.  One must solve for the vacuum
state in addition to the states with nonzero longitudinal
momentum and then subtract.  

The vacuum-bubble contributions
are proportional to $\delta(0)$ and require regularization,
with the regulator removed after the vacuum subtraction.
One such regulator is to give the delta function a 
nonzero width measured by a particular parameter to be
taken to zero~\cite{Tadpoles}.\footnote{In the context of perturbative
amplitudes, the regulator can be the radius of an integration
contour taken to infinity~\protect\cite{MannheimPLB}.}

In any case, the resurrection of vacuum bubbles within
nonperturbative LF calculations is what makes contact
with the early work of Chang and Ma~\cite{ChangMa} and
of Yan~\cite{Yan}, where zero-mode contributions restore
the equivalence between LF and covariant perturbation 
theory.\footnote{For more recent discussions of equivalence,
see \protect\cite{Mannheim,Polyzou}.}
The importance of this equivalence and its implications
were recently emphasized by Collins~\cite{Collins}, as
illustrated in a calculation of the Greens function
associated with a one-loop self-energy correction in a
two-dimensional $\phi^3$ theory.\footnote{See also the 
analysis in \protect\cite{Martinovic}.}

Zero modes also play a role in the design of numerical
methods for the solution of field-theoretic bound-state
problems.\footnote{For a discussion of calculations for
analogous modes in the context of a quartic oscillator,
see \protect\cite{Girgus}.}  Even if one excludes vacuum-to-vacuum 
transitions, these modes still enter as accumulation points for mildly
singular integrals~\cite{Tsujimaru}.  The use of discretization, as in 
the discretized light-cone quantization (DLCQ) method
of Pauli and Brodsky~\cite{PauliBrodsky}, places a
measurable burden on these endpoint contributions.
The trapezoidal approximation used implicitly in DLCQ
can then have a numerical error that has a worse-than-canonical
dependence on the resolution.  This can be corrected by
including in the DLCQ Hamiltonian effective interactions
that represent the effects of zero modes~\cite{DLCQZeroModes}
or overcome by computing at very high resolution~\cite{HiResVary}.

Another approach is to use basis functions that provide
a representation of the expected endpoint behavior.  One
such attempt~\cite{BCH} used a representation that was
not sufficiently singular at the endpoints but had the
advantage of the matrix elements being computable
analytically.  As seen in Table~\ref{tab:phi4crit}
the result differs from what is obtained by a
high-resolution DLCQ calculation~\cite{HiResVary}.  As
discussed below in the context of tadpoles and vacuum bubbles,
an improved endpoint behavior requires numerical evaluation
of the matrix elements.  Because they are multidimensional,
beyond the lowest Fock sector, and singular (though integrable),
this is best done with an adaptive Monte Carlo algorithm
such as is at the heart of the VEGAS implementation~\cite{vegas}.

The extension to include effective interactions in the DLCQ
Hamiltonian is based on a solution to a zero-mode constraint
equation that is perturbative in powers of the DLCQ resolution.
One can instead consider nonperturbative solutions, as 
proposed by Werner and Heinzl~\cite{WernerHeinzl} and 
by Bender and Pinsky~\cite{BenderPinsky}.  The zero-mode
creation and annihilation operators are determined from
nonlinear equations that are dependent on matrix elements
of operators for ordinary modes.  For two-dimensional $\phi^4$ theory
nontrivial solutions exist for a range of coupling strengths
and can be associated with spontaneous symmetry breaking,
with a different vacuum for each solution.  The behavior
as a function of the coupling can be used to identify a
critical value and a critical exponent.  However, the
critical exponent is found to be the mean-field value
rather than the known value for the theory's universality
class~\cite{Simon}, and there are subtle renormalization 
issues that need to be fully addressed to obtain a finite
critical coupling~\cite{WernerHeinzl}.

The solution of the constraint then allows the construction of
the full Hamiltonian with effective interactions representing
the effects of zero modes.  The construction is done for 
each solution of the constraint, which is the mechanism for 
symmetry breaking in this approach.  Each solution corresponds
to a different choice of vacuum on which the eigenstates
of the Hamiltonian are built.

The remainder of this review focuses on two key considerations.
Section~\ref{sec:DLCQ} looks at the role of constrained zero
modes in DLCQ.  Section~\ref{sec:bubbles} considers the inclusion
of vacuum to vacuum transitions in LF calculations.  A brief
summary, including suggestions for additional work, is given
in Sec.~\ref{sec:summary}.

\section{Zero modes in DLCQ}\label{sec:DLCQ}

In the context of DLCQ, zero-mode contributions arise from a
constraint and are not independently dynamical.  The constraint
is the spatial average of the Euler-Lagrange equation for the 
field~\cite{Maskawa,Wittman,PinskyvdSHiller}, which is exactly solvable
only in very simple cases.  Usually the constraint and zero modes
are neglected as contributing effects that disappear in the
large-volume limit.  However, methods exist to solve the 
constraint approximately,\footnote{To be consistent with the 
existing literature, we define the LF spatial coordinate
$x^-$ as $(t-z)/\sqrt{2}$ in this section.} both perturbatively~\cite{DLCQZeroModes}
and nonperturbatively~\cite{WernerHeinzl,PinskyvdSHiller}, studied
with the expectation that symmetry-breaking effects will survive
the large-volume limit.

The perturbative approach~\cite{DLCQZeroModes}\footnote{This is
to be distinguished from ordinary perturbation theory as an
expansion in the coupling, which is considered in \protect\cite{BenderPinsky}
and \protect\cite{WernerHeinzl}.} is structured
to solve the constraint as a power series in the reciprocal
of the DLCQ resolution $K$.  The reciprocal determines the grid
spacing for the discretization of momentum fractions $x_i\equiv p^+_i/P^+$
for constituents with momenta $p^+_i$ in a system with total
momentum $P^+$.  Integrals over $x_i$ are then approximated by
\be
\int_0^1 dx_i f(x_i) =\frac{1}{2K}f(0)+\frac{1}{K}\sum_{n=1}^{K-1}f(n/K)+\frac{1}{2K}f(1).
\ee
The first and last terms are the zero-mode contributions that are
usually neglected.\footnote{Although the last term is not explicitly
a zero-mode contribution, when $x_i=1$, the momentum fractions for
the other constituents must be zero.  This is due to the positivity
of LF momenta $p_i$.}  They can be restored, however, through
effective interactions in the Hamiltonian.  In particular, this
can be used to show~\cite{DLCQZeroModes} that cubic scalar theories
do indeed have their expected unbounded negative spectra~\cite{Baym,Gross}
in LF quantization, due to zero-mode contributions.  The expansion
in $1/K$ is truncated at the level consistent with the numerical
approximation made for the DLCQ Hamiltonian.  Keeping higher
orders in the constraint equation would be considered inconsistent,
from this point of view.

As an example, consider an LF Hamiltonian ${\cal H}=\frac12\mu^2\tilde\phi^2+V(\tilde\phi)$.
The DLCQ approach imposes periodic boundary conditions on $\tilde\phi$ in a
box $-L/2<x^-<L/2$ such that $\tilde\phi$ has the plane-wave expansion
\be
\tilde\phi=\sum_{n>0}\frac{1}{\sqrt{4\pi n}}\left[e^{ik_n^+x^-}a_n+e^{-ik_n^+x^-}a^\dagger_n\right]+\phi_0
\ee
with $k_n^+=2\pi/L$ and $\phi_0$ the constant part of $\tilde\phi$.
The operators $a_n$ and $a^\dagger_n$ obey the commutation relation
$[a_n,a^\dagger_m]=\delta_{nm}$.  The Euler-Lagrange equation
\be
(2\partial_+\partial_- + \mu^2)\tilde\phi=-V'(\tilde\phi)
\ee
constrains $\phi_0$ because integration over the box yields
\be \label{eq:constraint0}
-\mu^2\phi_0=\frac{1}{L}\int_{-L/2}^{L/2}V'(\phi+\phi_0)dx^-,
\ee
where $\phi\equiv\tilde\phi-\phi_0$.  This implicitly determines
the DLCQ zero mode in terms of the dynamical modes.

For a shifted free scalar $\tilde\phi\rightarrow\tilde\phi+v$,
with $V(\tilde\phi)=v\mu^2\tilde\phi+\frac12\mu^2v^2$,
the constraint is trivial.  It reduces to
\be \label{eq:constraint1}
-\mu^2\phi_0=\frac{1}{L}\int_{-L/2}^{L/2}v\mu^2dx^-=v\mu^2,
\ee
or $\phi_0=-v$.  Thus DLCQ does recover the correct result.

A more complicated situation is that of $\phi^4$ theory, even in
two dimensions.  We work with a rescaled Hamiltonian~\cite{Rozowsky}
\bea
\widetilde{\cal P}^-&\equiv&\frac{2\pi}{\mu^2L}\Pminus \\
 &=& \frac14a_0^2+\frac{g}{24}a_0^4+\frac12\Sigma_2+g\Sigma_4 -C\\
 &&+\frac{g}{24}\sum_{n\neq0}\frac{1}{|n|}(a_0^2a_na_{-n}+a_na_{-n}a_0^2
      +a_na_0^2a_{-n}+a_na_0a_{-n}a_0  \nonumber \\
    && \rule{0.5in}{0mm}+a_0a_na_0a_{_n}
      +a_0a_na_{-n}a_a-3a_0^2)  \nonumber \\
 &&+\frac{g}{24}\sum_{k,l,m\neq0}\frac{\delta_{k+l+m,0}}{\sqrt{|klm|}}
     (a_0a_ka_la_m+a_ka_0a_la_m+a_ka_la_0a_m+a_ka_la_ma_0), \nonumber
\eea
where $a_0=\sqrt{4\pi}\phi_0$, $a_{-n}=a^\dagger_n$,
$g=\lambda/8\pi\mu^2$,\footnote{This $g$ is proportional to the
reciprocal of the $g$ in \protect\cite{PinskyvdSHiller}.} and 
\be
\Sigma_n=\frac{1}{n!}\sum_{i_1\cdots i_n\neq0}\frac{\delta_{i_1+\cdots+i_n,0}}
     {\sqrt{|i_1\cdots i_n|}}:a_{i_1}\cdots a_{i_n}: .
\ee
A symmetric ordering has been chosen, with $a_0$ treated on 
an equal footing with the other $a_n$.  Constants are
removed by the normal ordering in $\Sigma_2$ and $\Sigma_4$
and by explicit subtraction of a constant $C$, chosen to
keep the vacuum expectation value of $\Pminus$ at zero.  

The constraint equation can be written as
\be
- a_0=\frac{g}{3}a_0^3+2g\Sigma_3+\frac{2g}{3}
          \left(a_0\Sigma_2+\Sigma_2 a_0\right)
  +\frac{g}{3}\sum_{n\neq0}\frac{1}{|n|}a_na_0a_{-n}
\ee
and invoking it can simplify the Hamiltonian to
\bea
\widetilde{\cal P}^-&=&\frac{1}{2K}\bar\Sigma_2+\frac{g}{K^2}\bar\Sigma_4
-\frac{g}{24}a_0^4
+\frac{g}{24K^{3/2}}\sum_{klm\neq0}\frac{\delta_{k+l+m,0}}
                                   {\sqrt{x_{|k|} x_{|l|} x_{|m|}}}
                   (a_k a_l a_0 a_m + a_k a_0 a_l a_m) \nonumber \\
&&+\frac{g}{24K}\sum_{n\neq0}\frac{1}{x_{|n|}}
                                 (a_na_0^2a_{-n}-a_0 a_n a_{-n}a_0),
\eea
with
\be \label{eq:barSigma}
\widetilde\Sigma_n=\frac{1}{n!}\sum_{i_1\cdots i_n\neq0}\frac{\delta_{i_1+\cdots+i_n,0}}
     {\sqrt{|x_1\cdots x_n|}}:a_{i_1}\cdots a_{i_n}:.
\ee
The resolution dependence of the momentum fractions $x_i=i/K$ is such that
$\Sigma_n=\widetilde{\Sigma}_n/K^{n/2}$.

The nonzero-mode parts of the Hamiltonian are of order $1/K^2$.  Zero-mode
corrections need to be kept to one higher order, at $1/K^3$, consistent with
order of endpoint contributions to the trapezoidal rule.  This requires
expansion of $a_0$ to order $1/K^{3/2}$.  The solution of the constraint
equation to this order is~\cite{DLCQZeroModes}
\be \label{eq:phi4soln}
a_0=-\frac{2g}{K^{3/2}}\widetilde\Sigma_3=- 2g\Sigma_3.
\ee
The Hamiltonian to this order is
\be \label{eq:newphi4hamiltonian}
\widetilde{\cal P}^-=\frac{1}{2K}\widetilde\Sigma_2+\frac{g}{K^2}\widetilde\Sigma_4
-\frac{g^2}{12K^3}\sum_{klm\neq0}\frac{\delta_{k+l+m,0}}{\sqrt{x_{|k|} x_{|l|} x_{|m|}}}
                   (a_k a_l \widetilde\Sigma_3 a_m + a_k \widetilde\Sigma_3 a_l a_m)+\mathcal{O}(1/K^4).
\ee
One could then proceed with a diagonalization program for this Hamiltonian 
in terms of a Fock basis that does not include zero modes.  

However, there are in addition zero-mode loop effects~\cite{DLCQZeroModes,Hellerman,Taniguchi}
not captured by the DLCQ analysis.  To see how the loop can arise, consider
the coupled equations for Fock-state wave functions $\psi_n$ in $\phi^4$ theory
as formulated in the continuum rather than DLCQ:
\bea  \label{eq:coupledeqns}
\lefteqn{\sum_{i=1}^n\frac{1}{x_i}\psi_n+\frac{g/3}{\sqrt{n(n-1)}}
\sum_{i\neq j\neq k}\frac{\psi_{n-2}(x_1,\ldots,x_i+x_j+x_k,\ldots,x_n)}
                {\sqrt{x_ix_jx_k(x_i+x_j+x_k)}} } &&    \\
   && + \frac{g}{3}\sqrt{(n+1)(n+2)}\sum_i
       \int \frac{dx'_1 dx'_2
        \psi_{n+2}(x_1,\ldots,x'_1,\ldots,x'_2,\ldots,x_i-x'_1-x'_2,\ldots,x_n)}
                          {\sqrt{x'_1 x'_2 x_i (x_i-x'_1-x'_2)}}
                 \nonumber \\
   && + \frac{g}{2} \sum_{i\neq j}
     \int \frac{dx' \psi_n(x_1,\ldots,x',\ldots,x_i+x_j-x',\ldots,x_n)}
          {\sqrt{x_i x_j x' (x_i+x_j-x')}}
                =(M^2/\mu^2)\psi_n .
     \nonumber
\eea
From the same equation with $n$ replaced by $n+2$, we can obtain contributions 
to $\psi_{n+2}$ from $\psi_n$.  When introduced in place of $\psi_{n+2}$ in
Eq.~(\ref{eq:coupledeqns}), there is a term that couples $\psi_n$ to itself
of the form~\cite{DLCQZeroModes}
\bea
\lefteqn{\left(\frac{g}{3}\right)^2\sum_i 
    \int \frac{dx'_1 dx'_2}{\sqrt{x'_1 x'_2 x_i (x_i-x'_1-x'_2)}}} && \\
&&\times    \frac{1}{M^2/\mu^2-\sum_{j\neq i}\frac{1}{x_j}
                     -\frac{1}{x'_1}-\frac{1}{x'_2}-\frac{1}{x_i-x'_1-x'_2}} \nonumber \\
&&\times\sum_{k \neq i}\frac{\psi_n(x_1,\ldots,x_i-x'_1-x'_2,\ldots,x_k+x'_1+x'_2,\ldots,x_n)}
                           {\sqrt{x'_1 x'_2 x_k (x'_1+x'_2+x_k)}}.  \nonumber
\eea
This term represents the emission of two particles by the $i$th constituent that are
then absorbed by the $k$th.  This is a loop contribution with intermediate momentum
fractions $x'_1$ and $x'_2$.  The integrals over these
intermediate variables include endpoints where one or
both can be zero.  If one intermediate variable is zero, this corresponds to a
contribution from the $1/K^3$ term in the Hamiltonian shown in Eq.~(\ref{eq:newphi4hamiltonian}),
but if two are (near) zero the integrand has an integrable singularity that makes a 
contribution of order $1/K$ rather than the nominal $1/K^2$ behavior of
a double endpoint in DLCQ.  This additional correction is not captured
by the above analysis of the constraint equation and an extra term is
needed in the Hamiltonian~\cite{Taniguchi}.

In the nonperturbative approach, one first solves
the constraint equation for the matrix elements of $a_0$ 
and then uses these matrix elements to construct a
matrix representation of the Hamiltonian, which is
then diagonalized.  The matrix elements are computed relative
to a Fock basis in the dynamical modes.  In solving the 
constraint equation above the critical coupling, the vacuum
expectation value $\langle0|a_0|0\rangle=\sqrt{4\pi}\langle0|\tilde\phi|0\rangle$
is found to be nonzero and can take either sign.  The
Hamiltonian then comes in two versions, one for each sign.
This is, of course, to be interpreted as symmetry breaking
in a theory that was originally symmetric with respect
to $\tilde\phi\rightarrow-\tilde\phi$.

The matrix representation is made finite by a cutoff in
the values of $K$.  The matrix representations of the 
constraint equation and the Hamiltonian are computed
by inserting decompositions of the identity between
individual operators, which imbeds matrix elements
of $a_0$ in the matrix representation of the 
constraint equation and creates a nonlinear
algebraic problem which was solved by an
iterative numerical procedure~\cite{PinskyvdSHiller}.

The approach encountered a serious difficulty with the
interpretation of the results in that the value of the 
critical coupling diverges with the cutoff in $K$ values.
Continuing the program requires some form of additional
renormalization; the critical coupling is known to have
a finite value~\cite{Simon}.  A procedure for doing this
is suggested in \cite{WernerHeinzl} and applied successfully
at low orders in particle number.  However, the critical
exponent for the low-order calculation is the mean-field
value of 1/2 rather than the known exact value of 1/8~\cite{Simon}.
Working to any finite order in particle number may not
be able to reproduce the 1/8 value.  

As an alternative to DLCQ, we next consider zero-mode contributions
without discretization and without constraint equations.  For
other approaches, see \cite{McCartor,Herrmann}.

\section{Vacuum Bubbles and Tadpoles}\label{sec:bubbles}

\subsection{Free scalar}\label{sec:free}

To begin, we consider a free scalar in two dimensions.
This will set the stage for consideration of the vacuum
in interacting theories, because the LF vacuum of even
a free theory is trivial only in a nontrivial way.

The Lagrangian is simply
\be
{\cal L}_{\rm free}=\frac12(\partial_\mu\phi)^2-\frac12\mu^2\phi^2,
\ee
where $\mu$ is the mass of the boson.
The light-front Hamiltonian density is
\be
{\cal H}_{\rm free}=\frac12 \mu^2 \phi^2.
\ee
We drop the $+$ superscript from the LF longitudinal
momenta, to simplify the notation, and write the
mode expansion for the field $\phi$ as\footnote{Contrary to
the previous section, to again be consistent with existing
literature, we use $x^-=t-z$ in this section.}
\be \label{eq:phi}
\phi(x^+=0,x^-)=\int \frac{dp}{\sqrt{4\pi p}}
   \left\{ a(p)e^{-ipx^-/2} + a^\dagger(p)e^{ipx^-/2}\right\}.
\ee
For the creation and annihilation operators, the nonzero commutation relation is
\be \label{eq:acomm}
[a(p),a^\dagger(p')]=\delta(p-p').
\ee
The free LF Hamiltonian is $\Pfree=\Pminus_{11}+\Pminus_{02}+\Pminus_{20}$
with 
\bea \label{eq:Pfree}
\Pminus_{11}&=&\int dp \frac{\mu^2}{p} a^\dagger(p)a(p), \\
\label{eq:Pminus02}
\Pminus_{02}&=&\frac{\mu^2}{2}\int \frac{dp_1 dp_2}{\sqrt{p_1p_2}}\delta(p_1+p_2)
                 a(p_1)a(p_2), \\
\label{eq:Pminus20}
\Pminus_{20}&=&\frac{\mu^2}{2}\int \frac{dp_1 dp_2}{\sqrt{p_1p_2}}\delta(p_1+p_2)
                 a^\dagger(p_1)a^\dagger(p_2).
\eea
The subscripts indicate the number of creation and annihilation operators.

We can then specify the eigenvalue problem for the free vacuum as
\be
\Pfree\vac=P_{\rm vac}^-\vac.
\ee
The first term in $\Pfree$ is just the standard LF Hamiltonian for a
free scalar, for which the vacuum eigenstate is the Fock vacuum $|0\rangle$.
If we treat the other two terms as a perturbation, we can write
\be
\Pfree=\Pminus_{11}+\lambda(\Pminus_{02}+\Pminus_{20}),
\ee
where $\lambda$ is to be set to 1 after identifying orders in 
the perturbation, as usual.  We then expand
\be
\vac=\sum_{n\,{\rm even}}\lambda^n|n\rangle,
\ee
with 
\be
|n\rangle=\int\prod_i^n dp_i\,\psi^{(n)}(p_i)\frac{1}{\sqrt{n!}}\prod_i^na^\dagger(p_i)|0\rangle,
\ee
and
\be
\Pvac=\sum_{n\,{\rm even}}\lambda^n P_{\rm vac}^{-(n)}.
\ee
By collecting equal powers of $\lambda$ we have, in general,
\be  \label{eq:generaln}
\Pminus_{11}|n\rangle+(\Pminus_{02}+\Pminus_{20})|n-2\rangle=\sum_{m=0}^n P_{\rm vac}^{-(m)}|n-m\rangle.
\ee
The specific case of $\lambda^0$ is $\Pminus_{11}|0\rangle=P_{\rm vac}^{-(0)}|0\rangle$,
which implies $P_{\rm vac}^{-(0)}=0$, as expected.  

For $\lambda^2$ we have
\be
\Pminus_{11}|2\rangle+(\Pminus_{02}+\Pminus_{20})|0\rangle=P_{\rm vac}^{-(2)}|0\rangle.
\ee
Projection onto $\langle0|$ yields
\be
\langle0|\Pminus_{02}+\Pminus_{20}|0\rangle=P_{\rm vac}^{-(2)},
\ee
and therefore $P_{\rm vac}^{-(2)}=0$, also.  This, combined with the
fact that $\Pminus_{02}|0\rangle=0$, leaves
\be
\Pminus_{11}|2\rangle+\Pminus_{20}|0\rangle=0.
\ee
Substitution of the expressions for $|2\rangle$ and the operators yields
\bea
\lefteqn{\int dp \frac{\mu^2}{p} a^\dagger(p)a(p)\int dp_1 dp_2\psi^{(2)}(p_1,p_2)
\frac{1}{\sqrt{2}}a^\dagger(p_1)a^\dagger(p_2)|0\rangle}&& \\
  && =-\frac{\mu^2}{2}\int \frac{dp_1 dp_2}{\sqrt{p_1p_2}}\delta(p_1+p_2)
                 a^\dagger(p_1)a^\dagger(p_2)|0\rangle. \nonumber
\eea
The action of the annihilation operator $a(p)$ reduces this to
\bea
\lefteqn{\int dp_1 dp_2\left(\frac{\mu^2}{p_1}+\frac{\mu^2}{p_2}\right)\psi^{(2)}(p_1,p_2)
\frac{1}{\sqrt{2}}a^\dagger(p_1)a^\dagger(p_2)|0\rangle}&& \\
 && =-\frac{\mu^2}{2}\int \frac{dp_1 dp_2}{\sqrt{p_1p_2}}\delta(p_1+p_2)
                 a^\dagger(p_1)a^\dagger(p_2)|0\rangle. \nonumber
\eea
Projection onto the two-body Fock state gives us
\be \label{eq:psi2}
\psi^{(2)}(p_1,p_2)=-\frac{1}{\sqrt{2}}\frac{\delta(p_1+p_2)}{\sqrt{p_1p_2}}\frac{1}{\frac{1}{p_1}+\frac{1}{p_2}}.
\ee

For $\lambda^4$, projection of (\ref{eq:generaln}) onto $\langle0|$ leaves
\be
\langle0|\Pminus_{11}|4\rangle+\langle0|(\Pminus_{02}+\Pminus_{20})|2\rangle=P_{\rm vac}^{-(4)}.
\ee
The first term is, of course, zero, and only $\Pminus_{02}$ contributes to the 
second term.  On substitution of the expressions for $|2\rangle$, including the
form of $\psi^{(2)}$, and for $\Pminus_{02}$, this can be reduced to
\be
P_{\rm vac}^{-(4)}=-\frac{\mu^2}{2}\int \frac{dp_1 dp_2}{p_1 p_2}\frac{\delta(p_1+p_2)^2}{\frac{1}{p_1}+\frac{1}{p_2}}.
\ee
The transformation used previously in the Introduction, namely $Q\equiv p_1+p_2$, 
$x\equiv p_1/Q$, and $dp_1 dp_2=QdQdx$, brings this to an explicit expression for 
the leading contribution to the LF vacuum energy
\be  \label{eq:Pvac4}
P_{\rm vac}^{-(4)}=-\frac{\mu^2}{2}\int_0^\infty dQ\delta(Q)^2\int_0^1 dx
=-\frac{\mu^2}{2}\delta(0)\int_0^\infty dQ\delta(Q)=-\frac{\mu^2}{4}\delta(0).
\ee

\begin{figure}[ht]
\vspace{0.2in}
\begin{center}
\includegraphics[width=5cm]{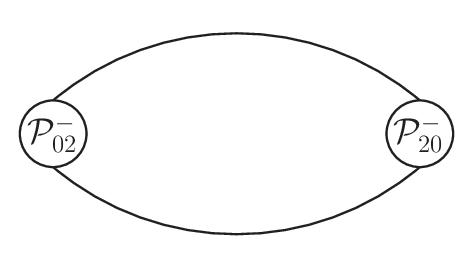}
\caption{\label{fig:FreeLoop}
Leading-order vacuum bubble for a free scalar
in light-front quantization that includes
interactions $\Pminus_{02}$ and $\Pminus_{20}$
defined in Eqs.~(\ref{eq:Pminus02}) and (\ref{eq:Pminus20}) of the text.
}
\end{center}
\end{figure}

This result is the contribution of the one-loop vacuum bubble, consistent with
the bubble computed by Collins~\cite{Collins}, with a graphical representation
given in Fig.~\ref{fig:FreeLoop}.  The remaining delta function
is simply the statement of momentum conservation in the bubble; this divergence
is not a serious problem in perturbation theory, because such contributions are
subtracted by hand.  For nonperturbative calculations, where bubbles are implicit,
this divergence needs to be regulated, which we do by replacing the delta functions
in $\Pminus_{02}$ and $\Pminus_{20}$ with a model function $\deps$ that has a width
parameter $\epsilon$.  This admits near-zero modes to the calculation, which
we call ephemeral modes~\cite{Tadpoles}, given that they are removed in a 
final limit of $\epsilon\rightarrow0$, once the vacuum is subtracted from physical states.

Various models can be considered for $\deps$, such as the following:
\be
\deps(p)=\frac{1}{2\epsilon}e^{-p/\epsilon}, \;\;
\deps(p)=\frac{1}{\epsilon\sqrt{\pi}}e^{-p^2/\epsilon^2},\;\;
\mbox{and} \;\;
\deps(p)=\frac{1}{2\epsilon}\theta(\epsilon-p).
\ee
Each is normalized such that $\int_0^\infty dp\,\deps(p)=\frac12$,
consistent with a limit to $\delta(p)$.  Many intermediate 
quantities will be model dependent but results for physical
states will not.  

The contribution of the one-loop vacuum bubble is proportional to $\int_0^\infty dQ\deps(Q)^2$,
as in Eq.~(\ref{eq:Pvac4}).
Given that $\delta(0)=L/4\pi$ in a finite one-dimensional LF volume $L$ in the $x^-$ 
direction,\footnote{In the limit $p\rightarrow0$, we 
have $\delta(p)=\int_{-\infty}^\infty\frac{dx^-}{4\pi}e^{ipx^-/2}\rightarrow\int_0^L\frac{dx^-}{4\pi}=\frac{L}{4\pi}$.}
setting $\int_0^\infty dQ\deps(Q)^2\rightarrow\frac12\delta(0)\rightarrow L/8\pi$ provides a direct link between the width
parameter $\epsilon$ and the volume scale $L$.  For the exponential, Gaussian,
and step models listed above, we have $\pi/L$, $\sqrt{8\pi}/L$, and $2\pi/L$,
respectively, for $\epsilon$ from direct computation of $\int_0^\infty dQ\deps(Q)^2$.

A nonperturbative solution for the vacuum can be found in the form of a 
Fock-state expansion in these ephemeral modes
\be  \label{eq:vac}
\vac=\sum_{n\,{\rm even}}\int\prod_i^ndp_i\deps(\sum^n p_i)
   \frac{c_n}{\sqrt{\prod_i^np_i}\sum_i^n\frac{1}{p_i}}
   \frac{1}{\sqrt{n!}}\prod_i^na^\dagger(p_i)|0\rangle.
\ee
The width of the delta-function replacement $\deps$ 
limits significant contributions to be near zero.
The specific momentum dependence included in place of $\psi^{(n)}$
is extrapolated from the 
perturbative solution for $\psi^{(2)}$ in Eq.~(\ref{eq:psi2})
and provides for finite contributions to the normalization.
The triviality of the free vacuum is recovered in the limit
that $\epsilon\rightarrow0$ and the ephemeral modes are removed.

The vacuum is normalized to 1:
\be
1=\langle{\rm vac}\vac
=\sum_n \int \prod_i^n dp_i\deps(\sum_i^np_i)^2\frac{|c_n|^2}{\prod_i^np_i\left(\sum_i^n\frac{1}{p_i}\right)^2}.
\ee
This can be simplified by the change in integration variables used previously, from
$p_i$ to $Q\equiv\sum_i^np_i$ and $x_i\equiv p_i/Q$, with the 
constraint that $\sum_i^n x_i=1$.  The Jacobian of the transformation
is again $Q^{n-1}$, and the vacuum normalization reduces to
\be
1=\int Q \deps(Q)^2dQ\sum_n\int\prod_i^n dx_i \delta(1-\sum_i^nx_i)
\frac{|c_n|^2}{\prod_i^nx_i\left(\sum_i^n \frac{1}{x_i}\right)^2}.
\ee
The leading integral over $Q$, which has been factored out, is
model dependent but finite.  The integrals over $x_i$ have mildly
singular integrands but are integrable, though not to a closed form; they are
best estimated by adaptive multidimensional Monte Carlo methods~\cite{vegas}.\footnote{Integrals
such as these are poorly approximated by a DLCQ approach because that is a multidimensional
trapezoidal approximation which is less efficient than Monte Carlo and because the 
singular behavior of the integrands is not well represented by equally
spaced points.}

To obtain equations for the coefficients $c_n$, we project
$\Pfree\vac=\Pvac\vac$ onto 
$\int \prod_i^m \frac{dq_i\,\deps(\sum_i^mq_i)}{\sqrt{\prod_i^m q_i}\sum_i^m\frac{1}{q_i}}
  \frac{1}{\sqrt{m!}}\prod_i^ma^\dagger(q_i)|0\rangle$, which yields
\bea
\lefteqn{\mu^2\int\frac{\prod_i^mdq_i\deps(\sum_i^m q_i)^2}{\left(\prod_i^mq_i\right)\left(\sum_i^m \frac{1}{q_i}\right)}c_m} && \\
&&+\frac{\mu^2}{2}\sqrt{m(m-1)}\int \frac{\prod_i^mdq_i\deps(\sum_i^m q_i)\deps(\sum_i^{m-2} q_i)\deps(q_{m-1}+q_m)}
   {\left(\prod_i^mq_i\right)\left(\sum_i^m \frac{1}{q_i}\right)\left(\sum_i^{m-2} \frac{1}{q_i}\right)}c_{m-2}  \nonumber \\
&&+\frac{\mu^2}{2}\sqrt{(m+2)(m+1)}\int\frac{\prod_i^mdq_i\deps(\sum_i^m q_i)}{\left(\prod_i^mq_i\right)\sum_i^m \frac{1}{q_i}} \nonumber \\
 && \rule{0.5in}{0mm} \times  \int\frac{dp'_1 dp'_2\deps(p'_1+p'_2)\deps(\sum_i^mq_i+p'_1+p'_2)}
             {p'_1p'_2\left(\sum_i^m\frac{1}{q_i}+\frac{1}{p'_1}+\frac{1}{p'_2}\right)}c_{m+2} \nonumber \\
&&=\Pvac\int\frac{\prod_i^mdq_i\deps(\sum_i^m q_i)^2}{\left(\prod_i^mq_i\right)\left(\sum_i^m \frac{1}{q_i}\right)^2}c_m. \nonumber
\eea
In each term we define $Q$ to be the most extensive sum over momenta; this
is $Q=\sum_i^m q_i$ in the first and second terms, as well as the 
right-hand side, and $Q=\sum_i^m q_i+p'_1+p'_2$ in the third term.  The 
individual momenta are then written in terms of fractions $x_i=q_i/Q$
and $x_{m+j}=p'_j/Q$.  With these new variables, the system of 
equations becomes
\bea  \label{eq:Qxeqn}
\lefteqn{\mu^2\int \deps(Q)^2dQ\int\frac{\prod_i^m dx_i\delta(1-\sum_i^m x_i)}{\left(\prod_i^mx_i\right)\left(\sum_i^m \frac{1}{x_i}\right)}c_m} && \\
&&+\frac{\mu^2}{2}\sqrt{m(m-1)}\int Q\deps(Q)dQ  \nonumber \\
&& \rule{0.5in}{0mm} \times \int \frac{\prod_i^mdx_i\delta(1-\sum_i^m x_i)\deps(Q\sum_i^{m-2} x_i)\deps(Q(x_{m-1}+x_m))}
   {\left(\prod_i^mx_i\right)\left(\sum_i^m \frac{1}{x_i}\right)\left(\sum_i^{m-2} \frac{1}{x_i}\right)}c_{m-2}  \nonumber \\
&&+\frac{\mu^2}{2}\sqrt{(m+2)(m+1)}\int Q\deps(Q)dQ  \nonumber \\
&& \rule{0.5in}{0mm} \times\int\frac{\prod_i^{m+2}dx_i\delta(1-\sum_i^{m+2} x_i)\deps(Q\sum_i^m x_i)\deps(Q(x_{m+1}+x_m))}
     {\left(\prod_i^{m+2}x_i\right)\left(\sum_i^m \frac{1}{x_i}\right)\left(\sum_i^{m+2} \frac{1}{x_i}\right)}c_{m+2} \nonumber \\
&&=\Pvac\int Q\deps(Q)^2dQ\int\frac{\prod_i^mdx_i\delta(1-\sum_i^m x_i)}{\left(\prod_i^mx_i\right)\left(\sum_i^m \frac{1}{x_i}\right)^2}c_m. \nonumber
\eea
The integrals over $Q$ are model dependent, but, if a model 
has $\epsilon$ as the only momentum scale, as is the case
for the models quoted above, the $\epsilon$ dependence can
be factored.  For any such model we can define the following
three $\epsilon$-independent quantities:
\bea \label{eq:abc}
\alpha&\equiv&\epsilon\int \deps(Q)^2 dQ, \;\;
\beta\equiv\int Q \deps(Q)^2 dQ, \\
\mbox{and}\;\; \gamma(y)&\equiv&\frac{\epsilon}{\alpha}\int Q \deps(Q) \deps(yQ)\deps((1-y)Q) dQ.
\nonumber
\eea
For the exponential, Gaussian, and step models listed above,
$\gamma(y)$ is 1/4, $1/[\pi\sqrt{2}(1-y+y^2)]$, and 1/4, respectively, with
$0<y<1$.

The scaling with $\epsilon$ is then clearly such that $\Pvac\propto1/\epsilon$
because each term on the left in Eq.~(\ref{eq:Qxeqn}) is proportional to $\alpha/\epsilon$.  We
can thus introduce a dimensionless eigenvalue $\eta$ via
\be
\eta\equiv\frac{\epsilon}{\mu^2}\frac{\beta}{\alpha}\Pvac
\ee
The system of equations then becomes
\bea
\lefteqn{\int\frac{\prod_i^mdx_i\delta(1-\sum_i^m x_i)}{\left(\prod_i^mx_i\right)\left(\sum_i^m \frac{1}{x_i}\right)}c_m} && \\
&&+\frac{1}{2}\sqrt{m(m-1)}\int \frac{\prod_i^mdx_i\delta(1-\sum_i^m x_i)\gamma(\sum_i^{m-2} x_i)}
   {\left(\prod_i^mx_i\right)\left(\sum_i^m \frac{1}{x_i}\right)\left(\sum_i^{m-2} \frac{1}{x_i}\right)}c_{m-2}  \nonumber \\
&&+\frac{1}{2}\sqrt{(m+2)(m+1)}\int\frac{\prod_i^{m+2}dx_i\delta(1-\sum_i^{m+2} x_i)\gamma(\sum_i^m x_i)}
     {\left(\prod_i^{m+2}x_i\right)\left(\sum_i^m \frac{1}{x_i}\right)\left(\sum_i^{m+2} \frac{1}{x_i}\right)}c_{m+2} \nonumber \\
&&=\eta\int\frac{\prod_i^mdx_i\delta(1-\sum_i^m x_i)}{\left(\prod_i^mx_i\right)\left(\sum_i^m \frac{1}{x_i}\right)^2}c_m, \nonumber
\eea
with the dependence on $\gamma$, defined in Eq.~(\ref{eq:abc}), the only remaining model dependence.
This of course implies that $\eta$ is model dependent; however,
model dependence in the vacuum energy is not a concern because it is
always subtracted.

As shown earlier, the momentum scale $\epsilon$ for the ephemeral modes is
inversely proportional to the one-dimensional volume $L$. For the integral
that defines $\alpha$ in Eq.~(\ref{eq:abc}), we find $\alpha=\epsilon L/8\pi$ or 
\be
\epsilon=8\pi\alpha/L
\ee
as a model-dependent definition of $\epsilon$ relative to a
common length scale.  This also shows that $\Pvac$ is proportional
to the one-dimensional volume and admits a finite vacuum energy
density $\Pvac/L$.

A state with nonzero momentum can be created from the vacuum with
a single creation operator
\be
|P\rangle=a^\dagger(P)\vac.
\ee
It is an eigenstate of $\Pfree$ with eigenvalue $\mu^2/P+\Pvac$, 
provided that $P\gg\epsilon$, because we can write
\bea
\Pfree a^\dagger(P)\vac&=&[\Pfree,a^\dagger(P)]\vac+a^\dagger(P)\Pfree\vac \\
   &=&\left[\frac{\mu^2}{P}a^\dagger(P)+\frac{\mu^2}{\sqrt{P}}\int \frac{dp}{\sqrt{p}}\deps(p+P)a(p)\right]\vac+\Pvac a^\dagger(P)\vac.
   \nonumber
\eea
The second term in the square bracket does not contribute, given the 
difference in scales between the physical and ephemeral modes, leaving
the expected eigenvalue.

\subsection{Shifted scalar}\label{sec:shifted}

If we introduce a shift in the field $\phi$ by an amount $v$,
the Lagrangian becomes 
\be
{\cal L}={\cal L}_{v=0}-\mu^2 v\phi-\frac12\mu^2 v^2,
\ee
and the Hamiltonian is $\Pminus=\Pfree+\Pint$, with
\be
\Pint=\int dx^- [\mu^2 v \phi+\frac12\mu^2v^2]
=\sqrt{4\pi}\mu^2 v \int \frac{dp}{\sqrt{p}}\deps(p)[a(p)+a^\dagger(p)]+\frac12\mu^2 v^2L.
\ee
The standard LF neglect of terms with only creation or annihilation operators
would mean that the impact of the shift is lost except for the constant term.  
We have instead introduced the regulated
delta function $\deps$ to retain all terms.

The lowest eigenstate of $\Pminus$ is the shifted vacuum state $\vac_v$:
\be
\left(\Pfree+\Pint\right)\vac_v=\Pvac\vac_v.
\ee
To incorporate this shift, we define~\cite{Tadpoles} an exponentiated operator
\be
B\equiv v\int dp \sqrt{4\pi p}\,\deps(p)[a(p)-a^\dagger(p)].
\ee
which shifts the field $\phi$
\be
e^B\phi(x^-)e^{-B}=\phi(x^-)+v.
\ee
When applied to the free Hamiltonian it generates the
interaction:
\be
e^B\Pfree e^{-B}=\int dx^- e^B\frac12\mu^2\phi^2e^{-B}=\Pfree+\Pint.
\ee
It also alters the free vacuum to the shifted vacuum with
\be
\vac_v=e^B\vac
\ee
and normalization
\be
_v\langle{\rm vac}\vac_v=\langle{\rm vac}|e^{B^\dagger}e^B\vac=\langle{\rm vac}|e^{-B}e^B\vac=\langle{\rm vac}\vac=1.
\ee

The shifted vacuum $\vac_v$ is the eigenstate because
\be
\left(\Pfree+\Pint\right)\vac_v=e^B\Pfree e^{-B}e^B\vac=e^B\Pfree\vac=\Pvac e^B\vac=\Pvac\vac_v.
\ee
The vacuum expectation value of the field is
\bea
_v\langle{\rm vac}|\phi(x^-)\vac_v&=&\langle{\rm vac}|e^{B^\dagger}\phi(x^-)e^B\vac
=\langle{\rm vac}|e^{-B}\phi(x^-)e^B\vac \nonumber \\
&=&\langle{\rm vac}|(\phi(x^-)-v)\vac=-v,
\eea
which restores the shift.

This shows that incorporation of ephemeral modes allows LF quantization
to replicate the known results for a shifted scalar field.\footnote{These 
results can also be obtained by considering the LF as a limit from ET
quantization~\protect\cite{Hornbostel,Ji,ETtoLF}.}  In the limit
that the regulator $\epsilon$ goes to zero, the ephemeral modes are removed but only
at the end of the calculation.

\subsection{$\phi^4$ theory}\label{sec:phi4}

The Lagrangian for two-dimensional $\phi^4$ theory is
\be
{\cal L}=\frac12(\partial_\mu\phi)^2-\frac12\mu^2\phi^2-\frac{\lambda}{4!}\phi^4,
\ee
where $\lambda$ is the coupling constant.
The LF Hamiltonian density is
\be
{\cal H}=\frac12 \mu^2 \phi^2+\frac{\lambda}{4!}\phi^4.
\ee
The LF Hamiltonian $\Pminus=\Pfree+\Pint$ then has several terms
arising from the interaction
$\Pint=\Pminus_{04}+\Pminus_{40}+\Pminus_{22}+\Pminus_{13}+\Pminus_{31}$
given by
\bea
\label{eq:Pminus04}
\Pminus_{04}&=&\frac{\lambda}{24}\int \frac{dp_1dp_2dp_3dp_4}
                              {4\pi \sqrt{p_1p_2p_3p_4}} 
     \deps(\sum_i^4 p_i) a(p_1)a(p_2)a(p_3)a(p_4), \\
\label{eq:Pminus40}
\Pminus_{40}&=&\frac{\lambda}{24}\int \frac{dp_1dp_2dp_3dp_4}
                              {4\pi \sqrt{p_1p_2p_3p_4}} 
      \deps(\sum_i^4 p_i)a^\dagger(p_1)a^\dagger(p_2)a^\dagger(p_3)a^\dagger(p_4), \\
\label{eq:Pminus22}
\Pminus_{22}&=&\frac{\lambda}{4}\int\frac{dp_1 dp_2}{4\pi\sqrt{p_1p_2}}
       \int\frac{dp'_1 dp'_2}{\sqrt{p'_1 p'_2}} 
       \delta(p_1 + p_2-p'_1-p'_2) \\
 && \rule{2in}{0mm} \times a^\dagger(p_1) a^\dagger(p_2) a(p'_1) a(p'_2),
   \nonumber \\
\label{eq:Pminus13}
\Pminus_{13}&=&\frac{\lambda}{6}\int \frac{dp_1dp_2dp_3}
                              {4\pi \sqrt{p_1p_2p_3(p_1+p_2+p_3)}} 
     a^\dagger(p_1+p_2+p_3)a(p_1)a(p_2)a(p_3), \\
\label{eq:Pminus31}
\Pminus_{31}&=&\frac{\lambda}{6}\int \frac{dp_1dp_2dp_3}
                              {4\pi \sqrt{p_1p_2p_3(p_1+p_2+p_3)}} 
      a^\dagger(p_1)a^\dagger(p_2)a^\dagger(p_3)a(p_1+p_2+p_3).
\eea
As in the free case, the subscripts represent the number of creation and annihilation operators
in each term.  Those interaction terms with only creation or only annihilation operators,
written as $\Pminus_{04}$ and $\Pminus_{40}$, are those dropped in
standard LF calculations and here have been regulated with $\deps$.

The vacuum $\vac_\lambda$ is the lowest eigenstate of $\Pminus$
with zero LF momentum: $\Pminus\vac_\lambda=\Pvac(\lambda)\vac_\lambda$.
The expansion in ephemeral modes takes the same form as in Eq.~(\ref{eq:vac}),
and the system of equations for the coefficients, derived in the 
same manner as in the free case, becomes
\bea  \label{eq:vacProblem}
\lefteqn{\int\frac{\prod_i^mdx_i\delta(1-\sum_i^m x_i)}{\left(\prod_i^mx_i\right)\left(\sum_i^m \frac{1}{x_i}\right)}c_m} && \\
&&+\frac{1}{2}\sqrt{m(m-1)}\int \frac{\prod_i^mdx_i\delta(1-\sum_i^m x_i)\gamma(\sum_i^{m-2} x_i)}
   {\left(\prod_i^mx_i\right)\left(\sum_i^m \frac{1}{x_i}\right)\left(\sum_i^{m-2} \frac{1}{x_i}\right)}c_{m-2}  \nonumber \\
&&+\frac{1}{2}\sqrt{(m+2)(m+1)}\int\frac{\prod_i^{m+2}dx_i\delta(1-\sum_i^{m+2} x_i)\gamma(\sum_i^m x_i)}
     {\left(\prod_i^{m+2}x_i\right)\left(\sum_i^m \frac{1}{x_i}\right)\left(\sum_i^{m+2} \frac{1}{x_i}\right)}c_{m+2} \nonumber \\
&&+\frac{g}{24}\sqrt{m(m-1)(m-2)(m-3)}\int \frac{\prod_i^mdx_i\delta(1-\sum_i^m x_i)\gamma(\sum_i^{m-4} x_i)}
   {\left(\prod_i^mx_i\right)\left(\sum_i^m \frac{1}{x_i}\right)\left(\sum_i^{m-4} \frac{1}{x_i}\right)}c_{m-4}  \nonumber \\
&&+\frac{g}{24}\sqrt{(m+4)(m+3)(m+2)(m+1)}\int\frac{\prod_i^{m+4}dx_i\delta(1-\sum_i^{m+4} x_i)\gamma(\sum_i^m x_i)}
     {\left(\prod_i^{m+4}x_i\right)\left(\sum_i^m \frac{1}{x_i}\right)\left(\sum_i^{m+4} \frac{1}{x_i}\right)}c_{m+4} \nonumber \\
&&+\frac{g}{4}m(m-1)\int \frac{\prod_i^mdx_i\delta(1-\sum_i^m x_i)dx'_1dx'_2\delta(x'_1+x'_2-x_{m-1}-x_m)}
   {\left(\prod_i^mx_i\right)x'_1x'_2\left(\sum_i^m \frac{1}{x_i}\right)
                        \left(\sum_i^{m-2} \frac{1}{x_i}+\frac{1}{x'_1}+\frac{1}{x'_2}\right)}c_m \nonumber \\
&&+\frac{g}{6}(m-2)\sqrt{m(m-1)}\int \frac{\prod_i^mdx_i\delta(1-\sum_i^m x_i)}
   {\left(\prod_i^mx_i\right)\left(\sum_{m-2}^mx_i\right)\left(\sum_i^m \frac{1}{x_i}\right)
                        \left(\sum_i^{m-3} \frac{1}{x_i}+\frac{1}{\sum_{m-2}^mx_i}\right)}c_{m-2} \nonumber \\
&&+\frac{g}{6}m\sqrt{(m+2)(m+1)}\int \frac{\prod_i^mdx_i\delta(1-\sum_i^m x_i)\prod_i^3dx'_i\delta(\sum_i^3x'_i-x_m)}
   {\left(\prod_i^mx_i\right)\left(\prod_i^3x'_i\right)\left(\sum_i^m \frac{1}{x_i}\right)
                        \left(\sum_i^{m-1} \frac{1}{x_i}+\sum_i^3\frac{1}{x'_i}\right)}c_{m+2} \nonumber \\
&&=\eta(\lambda)\int\frac{\prod_i^mdx_i\delta(1-\sum_i^m x_i)}{\left(\prod_i^mx_i\right)\left(\sum_i^m \frac{1}{x_i}\right)^2}c_m, \nonumber
\eea
with $g=\lambda/4\pi\mu^2$ as a dimensionless coupling constant.  The dimensionful 
eigenenergy is $\Pvac(\lambda)=\frac{\mu^2}{\epsilon}\frac{\alpha}{\beta}\eta(\lambda)$.
The quantities $\alpha$ and $\beta$ and the function $\gamma$ are defined as
before in Eq.~(\ref{eq:abc}).

\begin{figure}[hb]
\vspace{0.2in}
\begin{center}
\begin{tabular}{cc}
\includegraphics[width=6cm]{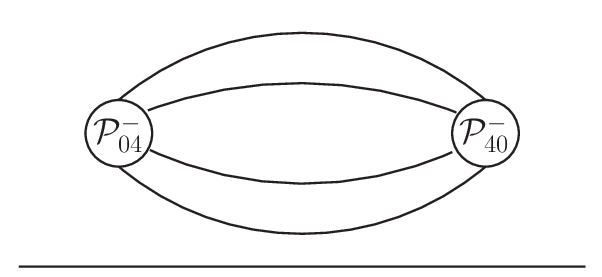} &
\includegraphics[width=6cm]{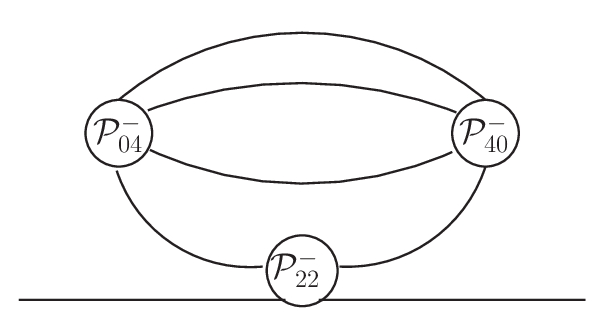} \\
(a) & (b) \\
\includegraphics[width=6cm]{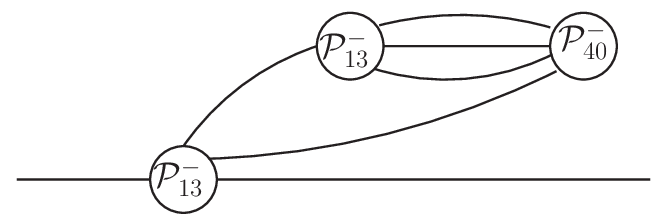} &
\includegraphics[width=6.5cm]{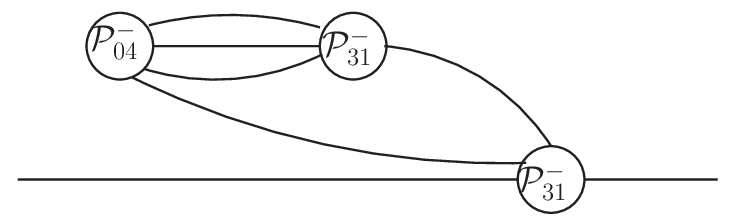} \\
(c) & (d)
\end{tabular}
\caption{\label{fig:phi4loops}
Lowest-order vacuum bubble (a) and tadpole graphs (b), (c), and (d) in $\phi^4$ theory. 
The vertices are labeled with the associated Hamiltonian
term.
}
\end{center}
\end{figure}

States with nonzero momentum are again built on this vacuum; however, with the interaction present they are of course
not as simple as the free case and instead have their own Fock state expansion
\be
|P\rangle=\sum_n \Psi_n^\dagger(P)\vac_\lambda,
\ee
with\footnote{The Fock-state wave functions $\psi_n$ introduced here are the standard LF wave functions
for $n$ constituents; however, the Fock states are built on the nontrivial vacuum which has its own
Fock-state wave functions $\psi^{(n)}$.  The momentum arguments of $\psi_n$ must total to $P$
whereas those of $\psi^{(n)}$ are all of order $\epsilon$.}
\be
\Psi_n^\dagger(P)\equiv \int \prod_i^n dp_i \delta(\sum_i^np_i-P)\psi_n(p_i)\frac{1}{\sqrt{n!}}\prod_i^n a^\dagger(p_i).
\ee
Such a state is to satisfy the eigenvalue problem
\be  \label{eq:PhysStateProblem}
\Pminus|P\rangle=\left[\frac{M^2}{P}+\Pvac(\lambda)\right]|P\rangle.
\ee
This will cause some mixing between physical and ephemeral modes as
the wave functions $\psi_n$ extend to zero momentum with a nominal 
behavior of
\be
\psi_n(p_i)\sim\frac{1}{\left(\sqrt{\prod_i^n p_i}\right)\left(\sum_i^n\frac{1}{p_i}\right)},
\ee
as discussed in the Introduction.
The mixing will introduce tadpole contributions that are absent from a
standard LF calculation.  There are also vacuum bubble contributions
when the physical and ephemeral modes do not mix, but these bubbles
contribute only to $\Pvac(\lambda)$ and are subtracted.  For a 
graphical representation of the lowest order contributions,
see Fig.~\ref{fig:phi4loops}.

These contributions can be seen explicitly in perturbation
theory~\cite{Tadpoles}.  If only perturbations due to 
$\Pminus_{04}$, $\Pminus_{40}$, and $\Pminus_{22}$ are kept, the equations for 
the lowest order Fock-state wave functions become
\bea
\label{eq:first}
\frac{\mu^2}{P}\psi_1&+&\frac{\lambda}{\sqrt{24}}\int \frac{\prod_i^4 dp_i}{4\pi\sqrt{\prod_i^4 p_i}}
\deps(\sum_i^4 p_i)\psi_5(p_1,\ldots,p_5)  \\
&& =\left(\frac{M^2}{P}+\Pvac\right)\psi_1, \nonumber \\
\label{eq:second}
\left(\sum_i^5\frac{\mu^2}{p_i}\right)\psi_5
&+&\frac{\lambda}{24}\frac{1}{5}\left[\frac{\deps(\sum_i^4 p_i)}{4\pi\sqrt{\prod_i^4 p_i}} 
                                       + (p_5\leftrightarrow p_1,p_2,p_3,p_4)\right]\psi_1 \\
&+&20\frac{\lambda}{4}\int \frac{dp'_1 dp'_2}{4\pi\sqrt{p_1p_2p'_1p'_2}}\delta(p_1+p_2-p'_1-p'_2)\psi_5(p'_1,p'_2,p_3,p_4,p_5)
=\frac{M^2}{P}\psi_5.  \nonumber
\eea
The second equation can be solved approximately by iteration of the
self interaction to first order in $\lambda$.
Substitution of the $\psi_5$ obtained by this approximation into the first equation
yields two contributions.  One corresponds to the bubble in Fig.~\ref{fig:phi4loops}(a)
and contributes the following to $\Pvac$~\cite{Tadpoles}:
\be
-\frac{\lambda^2}{\mu^2} \int \deps(Q)^2 dQ \int \frac{\prod_i^4 dx_i}{\prod_i^4 x_i}\delta(1-\sum_i^4 x_i),
\ee
which diverges as $1/\epsilon\sim L$.  The other corresponds to Fig.~\ref{fig:phi4loops}(b)
and contributes the following to $M^2/P$~\cite{Tadpoles}:
\be
\frac{\lambda^3}{P}\int \deps(Q)dQ\int \frac{\prod_i^4 dx_i}{(\prod_i^4 x_i)x_4\left(\sum_i^4 \frac{\mu^2}{x_i}\right)^2}.
\ee
This is finite and has the correct dependence on the total LF momentum for a self-energy 
correction.  Figures~\ref{fig:phi4loops}(c) and (d) also contribute when $\Pminus_{13}$
and $\Pminus_{31}$ are also kept.

The solutions of the vacuum problem stated in Eq.~(\ref{eq:vacProblem}) and the physical-state 
problem in Eq.~(\ref{eq:PhysStateProblem})
will provide for a finite mass value after the vacuum subtraction and for physical
Fock-state wave functions that can be used to compute expectation values.  This will
allow investigation of the critical coupling, where states with odd and even numbers
of constituents become degenerate and mix, which allows the field $\phi$ to have
a nonzero expectation value.

Another, related approach is that of the light-front coupled-cluster (LFCC) method~\cite{LFCC}.
Ephemeral modes can be included~\cite{LFCCzeromodes} by building the vacuum as
a generalized coherent state~\cite{Harindranath,Bartnik,coherentstates}.

\section{Summary}\label{sec:summary}

Zero modes have important roles to play in the structure
of light-front field theories and their solutions.  In
particular, the LF vacuum is not completely trivial.
Zero modes and the near-zero ephemeral modes act to induce
effects that are otherwise missing from LF calculations.
Although for many LF calculations these effects are not important,
complete consistency with equal-time calculations cannot
be achieved without zero modes.

With regard to the specifics of zero-mode effects, there are
various calculations still to be done, to fully understand
the mechanisms of zero modes and ephemeral modes.  For zero
modes within DLCQ, the perturbative approach to the constraint
equation and the subsequent construction of the effective
Hamiltonian, to include zero-mode loops, should be carried
out and used to study the impact of the added effective
interactions on the spectrum of two-dimensional $\phi^4$
theory and on its symmetry breaking.  Similarly, the
renormalization of the nonperturbative approach should be
systematized, so that many-body calculations can be attempted.

A better understanding of vacuum to vacuum transitions is
also needed.  Tadpoles clearly make an important contribution
to equivalence with equal-time calculations, as do 
vacuum bubbles.  Light-front calculations that include
them will be central to understanding symmetry breaking.
Two-dimensional $\phi^4$ theory is again an ideal model to consider.

Inclusion of tadpoles has already been shown to resolve a
disagreement with equal-time calculations~\cite{BCH}, but this was
not done from within a calculation that was exclusively light-front.
Instead, we have proposed~\cite{Tadpoles} a fundamental reinterpretation of
light-front Hamiltonians to include vacuum to vacuum
transitions in a way that calculations can be done strictly
within the light-front framework.

\backmatter

\bmhead{Acknowledgements}

This work was supported in part by 
the Minnesota Supercomputing Institute 
and the Research Computing and Data Services
at the University of Idaho through
grants of computing time.

\section*{Declarations}

\begin{itemize}
\item Funding: The work on zero modes in DLCQ by perturbation in the resolution~\cite{DLCQZeroModes} was funded
by the US Department of Energy, under Contract No.~DE-FG02-98ER41087.
\item Conflict of interest: Not applicable.
\item Ethics approval and consent to participate: Not applicable.
\item Consent for publication: Institutional consent not required.
\item Data availability: Not applicable. 
\item Materials availability: Not applicable.
\item Code availability: Not applicable.
\item Author contribution: Both authors contributed equally to the work, and read and agreed
to the manuscript submitted for publication.
\end{itemize}


\end{document}